New $^{34}$Cl proton-threshold states and the thermonuclear $^{33}$S(p,γ)$^{34}$Cl rate in ONe novae

A. Parikh[1,2,*], T. Faestermann[1,2], R. Hertenberger[2,3], R. Krücken[1,2], D. Schafstadler[1], H.-F. Wirth[2,3], T. Behrens[1,2], V. Bildstein[1,2], S. Bishop[1,2], K. Eppinger[1,2], C. Herlitzius[1,2], C. Hinke[1,2], M. Schlarb[1,2], D. Seiler[1,2], and K. Wimmer[1,2]

[1] Physik Department E12, Technische Universität München, D-85748 Garching, Germany

[2] Maier-Leibnitz-Laboratorium der Münchner Universitäten (MLL), D-85748 Garching, Germany

[3] Fakultät für Physik, Ludwig-Maximilians-Universität München, D-85748 Garching, Germany

ABSTRACT

Analysis of presolar grains in primitive meteorites has shown isotopic ratios largely characteristic of the conditions thought to prevail in various astrophysical environments. A possible indicator for a grain of ONe nova origin is a large $^{33}$S abundance: nucleosynthesis calculations predict as much as 150 times the solar abundance of $^{33}$S in the ejecta of nova explosions on massive ONe white dwarfs. This overproduction factor may, however, vary by factors of at least 0.01 – 3 because of uncertainties of several orders of magnitude in the $^{33}$S(p,γ)$^{34}$Cl reaction rate at nova peak temperatures ($T_{peak}$ ~ 0.1 – 0.4 GK). These uncertainties arise due to the lack of nuclear physics information for states within ~ 600 keV of the $^{33}$S+p threshold in $^{34}$Cl ($S_p(^{34}$Cl$) = 5143$ keV). To better constrain this rate we have measured, for the first time, the $^{34}$S($^3$He,t)$^{34}$Cl reaction over the region $E_x(^{34}$Cl$) = 4.9 - 6$ MeV. We confirm previous states and find 15 new states in this energy region. New $^{33}$S(p,$\gamma$)$^{34}$Cl resonances at $E_R = 281(2)$, $301(2)$ and $342(2)$ keV may dominate this rate at relevant nova temperatures. Our results could affect predictions of sulphur isotopic ratios in nova ejecta (e.g., $^{32}$S/$^{33}$S) that may be used as diagnostic tools for the nova paternity of grains.

---

[*] email: anuj.parikh@ph.tum.de

I. INTRODUCTION

Presolar grains embedded within primitive meteorites can provide information on stellar evolution and nucleosynthesis specific to different astrophysical phenomena. They are identified by their isotopic compositions, which may differ by orders of magnitude from that of the solar system. To date, carbonaceous (e.g., diamond, SiC, graphite), oxide (e.g., corundum, spinel), silicate, and silicon nitride grain types have been identified; ratios of carbon, nitrogen, oxygen, neon, magnesium and silicon isotopes, for example, have been measured. Most of the studied grains carry isotopic signatures thought to be indicative of origins in asymptotic giant branch (AGB) stars, giant stars or Type II supernovae (for reviews see e.g., [1-3]).

Recently, several grains exhibiting major-element isotopic ratios (e.g., $^{12}$C/$^{13}$C, $^{14}$N/$^{15}$N, $^{30}$Si/$^{28}$Si) possibly characteristic of classical nova explosions on ONe white dwarfs were discovered [4,5] (but see also [6]). A telltale sign of an ONe nova may be the observation of large (relative to solar) amounts of $^{33}$S (stable, 0.75%) in grains [4,7]; models of explosions on 1.35 M$_\odot$ ONe white dwarfs give an overproduction factor $X_{33}/X_{33_\odot} = 150$ for this isotope [8]. Indeed, "the predicted

$^{33}$S excess may provide a remarkable signature of a classical nova event" [7]. Sulphur has been observed in the ejecta of novae (e.g., Nova Aql 1982 [9,10]), and equilibrium condensation calculations predict the incorporation of sulphides into SiC grains [11]; however, only limited information on sulphur isotopic ratios is available from presolar grains (e.g., [7,12,13]). Sulphur measurements are complicated by, among other factors, contamination introduced in the chemical separation of SiC grains; despite these difficulties, the search for sulphur isotopic signatures in presolar grains continues [14,15].

Nova explosions are thought to occur with peak temperatures $T_{peak}$ = 0.1 – 0.4 GK, but the creation of elements above silicon in these proton-rich environments requires T > 0.3 GK (and hence, massive ONe white dwarfs) due to the large Coulomb barriers involved [8]. The most important destruction mechanism for $^{33}$S in nova explosions is the $^{33}$S(p,γ)$^{34}$Cl reaction (Q = 5142.75(12) keV [16]). This rate has been calculated using both the Hauser-Feshbach statistical model (e.g., [17,18]) and available experimental data [8]. Note that $^{34}$Cl has an isomeric state located at $E_x$ = 146 keV; we will denote the ground state (0+; $t_{1/2}$ = 1.53 s) as $^{34g}$Cl and the isomeric state (3+; $t_{1/2}$ = 32 m) as $^{34m}$Cl. The general nucleus will be written simply as $^{34}$Cl.

Iliadis et al. (2002) examined the effect of varying the total $^{33}$S(p,γ)$^{34}$Cl rate (as found using a statistical model calculation) in various ONe nova models [19]. They found reductions in the $^{33}$S nova abundance by factors of ~ 1000 when the rate was multiplied by 100, and enhancements by factors of ~3 when this rate was divided by 100 [19]. José et al. (2001) calculated the $^{33}$S(p,γ) rate using available experimental data for the resonance energies [20], along with calculated strengths for known threshold states (as no experimental measurements of these strengths exist, see below) [8]. They determined both the $^{33}$S(p,γ)$^{34g}$Cl and $^{33}$S(p,γ)$^{34m}$Cl rates at nova temperatures using information about the γ-decay branching ratios for the important resonances [20]. To reflect the uncertainty in their calculated strengths, they included an arbitrary overall factor f = 0 – 1 in their rates. If f = 1, their total rate for the $^{33}$S(p,γ)$^{34}$Cl reaction agrees with that used in Iliadis et al. (2002) to a factor of ~4 over nova temperatures. As the resonance strengths of José et al. (2001) were found by scaling calculated upper limits by 0.1, a more conservative range for f might be 0 – 10. This would imply that the $^{33}$S(p,γ)$^{34}$Cl rate calculated using currently available *experimental* information leads to variations by factors of at least ~0.01 – 3 (i.e. $X_{33}/X_{33\odot}$ ~ 1.5 - 450) in the amount of $^{33}$S produced in nova explosions, using results from the sensitivity study of Iliadis et al. (2002). We note that isotopic ratios (e.g., $^{32}$S/$^{33}$S), rather than overproduction factors, are the actual measured quantities in presolar grains. In this context, present uncertainties in the $^{33}$S(p,γ) rate lead to predicted $^{32}$S/$^{33}$S isotopic ratios from ~ 30 – 9700 (based upon the predicted nova value of $^{32}$S/$^{33}$S ~ 97 from the 1.35 M$_\odot$ ONe white dwarf model of [8]). Although the overall impact on the nova ejecta may be reduced because of convection and averaging within the upper layers of the envelope (neither of which were considered in the one-zone post-processing study of Iliadis et al. (2002)), the existence and impact of new resonances, which may significantly enhance the $^{33}$S(p,γ) rate, must also be explored. Given the typical uncertainty of ~ 5% in isotopic ratio determinations (e.g., [4]), experimental information on proton-threshold resonances in $^{33}$S(p,γ)$^{34}$Cl is desired to reduce the nuclear physics uncertainty in the amount of $^{33}$S produced in nova explosions.

The $^{33}$S(p,γ)$^{34}$Cl reaction may also play a role in γ-ray line astronomy. The potential observation of γ-rays emitted following the β+ decay of $^{34m}$Cl (E$_\gamma$ = 1.177, 2.128, 3.304 MeV, as well as from pair-annihilation) has been examined as a possible signature of a classical nova [21,22]. The short half-life (32 m) of $^{34m}$Cl (55.4% β+ to excited states of $^{34}$S, 44.6% γ-decay to $^{34g}$Cl) requires the ambient medium about an astrophysical event to clear rapidly enough following the nucleosynthesis of $^{34m}$Cl to permit the escape of γ-rays. Novae may provide a suitable environment [21], though current models predict the envelope to remain optically thick for ~ hours to days following $T_{peak}$ [23 – 25], implying significant suppression of β-delayed γ-rays from $^{34m}$Cl. Moreover, the maximum visual luminosity of a nova occurs ~days after $T_{peak}$, so that only a serendipitous observation and/or a large field-of-view instrument (such as SWIFT/BAT [26]) would likely be able to detect any early emission (i.e., occurring before the nova is discovered visually). Coc et al. (1999) consider thermally-induced transitions between $^{34m}$Cl and $^{34g}$Cl in an astrophysical plasma, and find the effective half-life of $^{34m}$Cl drops from the laboratory value of 32 m down to ~ 1 s over T = 0.1 – 0.4 GK [22]. José et al. (2001) adopt this effective lifetime in their model of a nova explosion on a 1.35 $M_\odot$ ONe white dwarf, and find a mean mass fraction X($^{34m}$Cl) = 7 x $10^{-7}$ in the ejecta, 30 minutes after $T_{peak}$ = 0.33 GK (at which most of the $^{34}$Cl was produced) [8]. This might be compared to the amount of the β-delayed γ-emitter $^{22}$Na ($t_{1/2}$ = 2.6 y; $E_\gamma$ = 1.275 MeV) expected in the ejecta from a nova explosion on a 1.35 $M_\odot$ ONe white dwarf: X($^{22}$Na) ~ $10^{-3}$ [27]. Given that 1.275 MeV γ-rays from the decay of $^{22}$Na have not yet been unambiguously observed in the Galaxy (but see [28]), as well as the above considerations on the short duration and early appearance of the emission from $^{34m}$Cl, the detection of γ-rays from the decay of $^{34m}$Cl would seem to be extremely challenging. In nova explosions, $^{34m}$Cl is formed only through the $^{33}$S(p,γ)$^{34m}$Cl reaction since $^{34}$Ar does not populate the isomer in its β-decay [20]. A dramatic increase in the $^{33}$S(p,γ) rate, through the discovery of new resonances and/or measurements of the important resonance strengths, would help to improve prospects for the detection of γ-rays from the decay of $^{34m}$Cl produced in nova explosions.

In a stellar environment, the resonant $^{33}$S(p,γ)$^{34}$Cl reaction rate is calculated as (in $cm^3 s^{-1} mol^{-1}$) [29]

$$N_A <\sigma v> = 1.5399 \times 10^{11} (\mu T_9)^{-3/2} \sum_i (\omega\gamma)_i \exp(-11.605 E_{R,i}/T_9) \qquad (1)$$

where $T_9$ is the temperature in GK, $\mu$ is the reduced mass of the $^{33}$S+p system in u, and the $E_{R,i}$ are the center-of-mass resonance energies in MeV. The resonance strengths $(\omega\gamma)_i$ (in MeV in eq. (1)) can be expressed as

$$(\omega\gamma) = \frac{2J_R+1}{(2J_p+1)(2J_t+1)} \frac{\Gamma_p \Gamma_\gamma}{\Gamma} \qquad (2)$$

where $J_R$, $J_p$ (=1/2) and $J_t$ (=3/2) are the spins of the resonance in $^{34}$Cl, the proton and $^{33}$S (ground-state) respectively. The total width of a resonance $\Gamma$ can be expressed as the sum of proton and γ-ray partial widths ($\Gamma_p$ and $\Gamma_\gamma$, respectively) for $^{34}$Cl states relevant to nucleosynthesis in ONe novae. The sum in eq. (1) allows for the contributions of all resonant states through which the reaction may proceed at the temperature of interest. Note that eq. (1) is valid only if these states are narrow ($\Gamma << E_R$) and isolated ($E_{R,i} - E_{R,j} >> \Gamma_i$). Since the proton-capture rates to both $^{34g}$Cl and $^{34m}$Cl are desired, each term in eq. (1) must also be multiplied by a factor $G_i$ (or 1-$G_i$) to account for the population of the ground and/or isomeric states in the γ-decay of excited states in $^{34}$Cl.

The most recent direct studies of resonances in the $^{33}$S(p,γ)$^{34}$Cl reaction were made by Dassie et al. (1977) (who examined resonances from $E_x$ = 6.13 – 7.08 MeV [30,31]) and Waanders et al. (1983) (who examined resonances from $E_x$ = 5.57 – 6.32 MeV [32]). Resonance energies, strengths, $J^\pi$ values and γ-decay schemes were found in both studies. Based on the γ-ray spectra, these two groups also deduced three states in $^{34}$Cl between the $^{33}$S+p threshold and $E_x$ = 5.576

MeV (the lowest-energy resonance that has been directly observed through $^{33}$S(p,γ)), at $E_x$ = 5.17, 5.39 and 5.54 MeV. Baumann et al. (1978) and van der Poel et al. (1982) observed a level at $E_x$ = 5.315 MeV through measurements of the ($^{12}$C,αnγ) and ($^{12}$C,pnγ) reactions, respectively [33,34]. Gamma-ray decay branching ratios for the resonances observed at $E_x$ > 5.57 MeV, as well as the states at $E_x$ = 5387 and 5541 keV have been measured by Waanders et al. (1983) [32]. In addition, Dassie at al. (1977) and van der Poel et al. (1982) measured the γ-decay schemes of the states at $E_x$ = 5172 and 5315 keV, respectively [30,31,34]. These past measurements were compiled in [20] and are presented in Table I.

Calculation of the thermonuclear resonant $^{33}$S(p,γ)$^{34}$Cl rate over nova $T_{peak}$ ~ 0.1 – 0.4 GK requires knowledge of resonance energies $E_R$ and strengths (ωγ) for states in $^{34}$Cl within ~600 keV of the proton-threshold ($S_p$($^{34}$Cl) = 5143 keV). There is no experimental information available for the strengths of known states below $E_x$ = 5.576 MeV. The $J^π$ values of table I indicate that these states would not be s or p-wave resonances. Proton and/or γ-ray partial widths $Γ_p$, $Γ_γ$ have not been measured for these states; proton spectroscopic factors are also unknown. As $^{34}$Cl is self-conjugate, no additional information can be obtained by appealing to the characteristics of a mirror nucleus. Given the exponential dependence of the reaction rate on the resonance energies, these quantities in particular must be known accurately. To both confirm past proton-threshold states and search for new states, we have measured, for the first time, the $^{34}$S($^3$He,t)$^{34}$Cl reaction over the relevant energy region for the $^{33}$S(p,γ)$^{34}$Cl rate in ONe novae.

## II. EXPERIMENT

The $^{34}$S($^3$He,t)$^{34}$Cl reaction was measured at the Maier-Leibnitz-Laboratorium (MLL) in Garching, Germany. Targets of Ag$_2$S (50 μg/cm$^2$, enriched to 99.999% in $^{34}$S) were prepared on to 10 μg/cm$^2$ carbon foils according to the method described in [35]; an Al foil target (100 μg/cm$^2$) was also used for calibration of the focal plane. An intense beam of $^3$He$^{2+}$ (I ~ 0.6 μA, over a 2.5 day period) from the ECR-like ion source [36] was brought to 25 MeV with the MP tandem accelerator. Light reaction products were momentum-analyzed in the Q3D magnetic spectrograph, with measurements at spectrograph angles of $\theta_{lab}$=15° and 25°. The rectangular aperture of the Q3D was fully open, giving an angular acceptance of 13.9 msr. The focal plane detector consisted of a multiwire gas proportional counter backed by a plastic scintillator [37,38], allowing for particle identification through determination of particle position, energy loss and residual energy. The observed energy resolution for the $^{34}$S($^3$He,t) spectra was about 10 keV FWHM, which is consistent with the intrinsic energy resolution of the Q3D ($\Delta E/E$ ~ 2 x 10$^{-4}$, contributing $\Delta E \approx$ 4 keV here) and SRIM calculations of energy loss in the Ag$_2$S target [39].

For these experimental conditions, tritons from the potential contaminant reactions $^{12}$C($^3$He,t), $^{16}$O($^3$He,t) and $^{32}$S($^3$He,t) were excluded from the detector by virtue of the Q-values of the respective reactions. Background from two broad states of the $^{13}$C($^3$He,t)$^{13}$N reaction (E$_x$ ($^{13}$N)= 7.9 and 8.9 MeV, Γ = 1500 and 230 keV, respectively [40]) was minimized through the use of foils enriched in $^{12}$C with our Ag$_2$S targets. A diffuse background from ($^3$He,t) reactions on Ag isotopes (populating the region E$_x$ ~ 9 – 11 MeV in the residual Cd nuclei) was observed; this was found to be featureless (see fig. 1) through measurements with a natural Ag target (50 μg/cm$^2$, on a 10 μg/cm$^2$ carbon backing). Negligible contamination due to peaks from ($^3$He,t) reactions on $^{33}$S and $^{36}$S isotopes (0.75% and 0.01% naturally, respectively) was expected based on the highly enriched nature of our target as well as measurements with a $Ag_2S$ target prepared using *natural* sulphur. This was confirmed through observation of no significant kinematic shifts for observed $^{34}Cl$ states between spectra at the different Q3D angles used for measurements. For example, for the most difficult case of distinguishing between tritons from the $^{34}S(^3He,t)$ and $^{33}S(^3He,t)$ reactions, the relative shift expected between $\theta_{lab}$=15° and 25° was ~ 6 keV.

Triton spectra were analyzed through least-squares fits of either multiple Gaussian or multiple exponentially-modified Gaussian (to account for a low-energy tail) functions. Peak widths were constrained to ≈10 – 15 keV FWHM based on fits of high-intensity, well-resolved peaks in the spectra. Background was assumed to be either constant or linear (increasing slowly with $E_x$). Peak centroids were thereby extracted. Fitting the data in these different ways gave consistent excitation energies within our stated systematic error (see below).

The focal plane was calibrated using 14 well-resolved, known states in $^{27}Si$ populated by the $^{27}Al(^3He,t)$ reaction, with 5.49 < $E_x(^{27}Si)$ < 6.63 MeV, and $\Delta E_x$ ~ 4 keV [20]. Second-degree polynomial least-squares fits of triton radius ρ to focal-plane position ($\chi_v^2$ ~ 1.2) were derived with this information for each spectrograph angle. These fits were then used to determine excitation energies for states in $^{34}Cl$ arising from the $^{34}S(^3He,t)$ reaction, at both $\theta_{lab}$ = 15° and 25°. Fig. 1 shows $^{34}S(^3He,t)^{34}Cl$ triton spectra at both measured angles, and Table I lists our derived averaged excitation energies for states in $^{34}Cl$. An overall uncertainty of ± 2 keV due to statistics and reproducibility is given in Table I for energies extracted in the present work. In addition to this, we find a systematic uncertainty of ± 3 keV due primarily to the beam energy distribution ($\Delta E_{beam}$ ≈ 1 part in 10 000) and uncertainty in the relative $Ag_2S$ and Al target thickness (each target thickness is known to ~ 10%)[1]. Negligible uncertainty is introduced from the relative Q-values of the $^{27}$Al($^3$He,t) and $^{34}$S($^3$He,t) reactions since all masses are known to better than 0.2 keV [16]. Table I also lists tentative states from our measurements which were observed with low statistics and/or at only one angle.

From Table I, it is evident that we have observed (albeit with low statistics in some cases) all previously known levels in $^{34}$Cl with 4.94 < $E_x$ < 5.94 MeV except for the known level at $E_x$ = 5315.0(3) keV. This is perhaps not unreasonable given the high spin (7$^+$) of this state. Our energies for these known states are in agreement with the compilation of Endt (1990) [20] for all levels but those at 5541 and 5576 keV, for which marginal disagreement is observed. In addition, we find 15 new states in $^{34}$Cl, with 9 of these new states within 600 keV of the proton threshold ($E_x$ = 5143 keV). Another state at $E_x$ = 5705 keV was strongly populated at 15°, but could not be confirmed at 25°. We have also significantly reduced the uncertainties in two states which had been previously been identified at 5010(13) and 4971(11) keV.

III. DISCUSSION

Using our new results for the level structure of $^{34}$Cl above the $^{33}$S+p threshold, we have determined a new thermonuclear $^{33}$S(p,γ)$^{34}$Cl reaction rate over typical nova $T_{peak}$ ~ 0.1 – 0.4 GK. Since γ-ray decay branching ratios were not measured in the present work, we restrict ourselves here to calculating the total $^{33}$S(p,γ) rate (as opposed to separate rates leading to $^{34g}$Cl and $^{34m}$Cl). The direct-capture component of the rate was estimated with $S_{DC}(0) \sim 30$ keV-b (see [18,29] for details). The resonant component was calculated according to eq. (1).

---

[1] We note that a recent experiment by Lotay et al. has determined energies 7.46 < $E_x$($^{27}$Si) < 7.84 to high-precision (~0.5 keV) [41]. If we use preliminary results from that experiment for $^{27}$Si energies within the region of concern for our calibration [42] instead of the values presented in [20], we obtain $^{34}$Cl energies in agreement with the values in Table I, within our stated systematic error. Uncertainties in our $E_x$($^{34}$Cl) are not affected by these high-precision measurements as uncertainty in the calibration, even with the energies from [20], contributes only $\Delta E_x$($^{34}$Cl)≈ 0.3 keV.

Resonance energies were calculated using a weighted average of energies from the present and past work (when possible) along with Q($^{33}$S(p,γ)$^{34}$Cl) = 5142.75(12) keV [16]. For a *lower limit* on the reaction rate we used all states with $E_x(^{34}Cl) < 6$ MeV ($E_R < 860$ keV – the Gamow window at T = 0.4 GK extends up to $E_R \sim 550$ keV) for which resonance strengths have been *measured* (see Table I). This lower limit is plotted as the lower dashed grey line in Fig. 2a. For an *upper limit* on the rate, we have added to the lower limit the contributions of those states with unknown strengths – i.e., states from the present work and previously-known states with unknown strengths ($E_R < 860$ keV). We have calculated these strengths using eq. (2) with the assumption $\Gamma_p << \Gamma_\gamma$. Proton widths $\Gamma_p$ were then calculated with (e.g. [18])

$$\Gamma_p = \frac{2\hbar^2}{\mu a^2} P_l C^2 S \theta^2_{s.p.}, \qquad (3)$$

where μ is the reduced mass of the $^{33}$S+p system, a is the interaction radius (= 1.25 x ($1^{1/3} + 33^{1/3}$) fm), $P_l$ is the penetrability of the Coulomb and centrifugal barrier for orbital angular momentum l (calculated using [43] for appropriate Coulomb wavefunctions), C is an isospin Clebsch-Gordan coefficient, S is a spectroscopic factor, and $\theta^2_{s.p.}$ is the single-particle reduced width (calculated using [44]). For the calculation of the upper limit, we have assumed all new states are s-wave resonances; for previously-known states with unknown strengths, we have assumed the lowest l-transfer consistent with any available spin-constraints. $C^2S$ factors were assumed as unity for all states with unknown strengths. Resonance parameters for all states used in the calculation of the upper limit are given in Tables I and II; the upper limit is plotted as the upper dashed grey line in Fig. 2a. Finally, our suggested new $^{33}$S(p,γ)$^{34}$Cl rate was determined as the geometric mean of our upper and lower limits (see the solid black line in Fig. 2a). The ratio between the upper and lower limits ranges from 6 x 10$^7$ at T = 0.1 GK to 50 at 0.4 GK. Over T = 0.01 – 0.4 GK, the direct-capture component is only relevant for the lower limit, where it dominates below T = 0.11 GK. The contribution of the level at $E_x$($^{34}$Cl) = 5315 keV (not observed in the present experiment) to the rate over nova $T_{peak}$ is negligible because of its 7+ character (requiring l=6 proton capture).

Our principal assumptions in calculating the upper limit of the $^{33}$S(p,γ) rate, namely maximal spectroscopic factors and l=0 character for all new states (as well as $\Gamma_p << \Gamma_\gamma$) lead to a very conservative estimate. This is highlighted through comparison of measured and calculated resonance strengths for $E_R$ > 400 keV (Tables I and II). Information from T=1 analogue states in $^{34}$S and $^{34}$Ar is potentially useful to constrain individual and/or the distribution of $J^\pi$ values (and hence the l-transfer) for some of the new resonances in $^{34}$Cl. However, with only 7 expected T=1 analogue states in the region 5.010 < $E_x$($^{34}$Cl) < 6.180 MeV [45] and 33 $^{34}$Cl states now known in this energy region (only one of which has been identified as T=1), this information is of limited use. We note that our simple treatment for the upper limit of the rate may certainly affect our suggested rate through its influence in the geometric mean.

Figure 2b shows our suggested $^{33}$S(p,γ)$^{34}$Cl rate over nova $T_{peak}$ as compared to the rates given in Iliadis et al. (2001) [18] and José et al. (2001) [8]. It is difficult to compare these rates directly since that from Iliadis et al. is based upon a statistical model calculation [17], and that from José et al. uses previously-known resonances [20] with calculated strengths when necessary. Our rate at T = 0.3 GK (at which the $^{33}$S(p,γ)$^{34}$Cl reaction begins to play a role in the abundance flow for novae on massive ONe white dwarfs [8]) is seven times larger than that of José et al. and 26 times larger than that of Iliadis et al. At T~0.3 GK, the upper limit in our rate calculation is dominated by the contributions of the three new resonances at $E_R$ = 281, 301 and 342 keV ($E_x$($^{34}$Cl) = 5424, 5444 and 5485 keV respectively); the lower limit is dominated by the contribution of the resonance at $E_R$ = 432 keV ($E_x$ = 5572 keV, the lowest energy resonance used in the lower limit calculation).

We can roughly estimate the impact of our new rate on nova nucleosynthesis using the calculations presented in Iliadis et al. (2002) [19] and José et al. (2001) [8]. In Iliadis et al. (2002), the $^{33}$S(p,γ) rate of Iliadis et al. (2001) was varied by factors of up to 100 in post-processing calculations using thermodynamic histories from nova explosions on 1.15 – 1.35 M$_\odot$ ONe white dwarfs, with $T_{peak}$ = 0.23 – 0.42 GK. In José et al. (2001), nova nucleosynthesis results from a hydrodynamic calculation on a 1.35 M$_\odot$ ONe white dwarf were provided: for example, $^{33}$S/$^{33}$S$_\odot$ ~ 150 ($^{32}$S/$^{33}$S ~ 97), and X($^{34m}$Cl)~ 7x10$^{-7}$, 30 min after $T_{peak}$ = 0.33 GK. If we use the relative differences between our new rate and those used in the studies of Iliadis et al. (2002) and José et al. (2001), we find a reduction in the $^{33}$S abundance by a factor of ~12 with our new rate, relative to the José et al. results. This would certainly significantly change the nature of any presolar grain nova signature to be expected from the observation of sulphur isotopic ratios. As separate rates for the $^{33}$S(p,γ)$^{34g}$Cl and $^{33}$S(p,γ)$^{34m}$Cl reactions must be provided to examine $^{34m}$Cl production in novae (which was not done in Iliadis et al. (2002), nor here), it is difficult to examine the effect of our new rate on $^{34m}$Cl production. The effect on $^{34}$S, the stable daughter of $^{34}$Cl, can however be estimated as above. With our new $^{33}$S(p,γ) rate we find an enhancement in $^{34}$S by only ~ 20% relative to the José et al. results, which is not encouraging for the potential detection of β-delayed γ-rays from $^{34m}$Cl. Of course, reliable implications of our new $^{33}$S(p,γ) rate should be examined through more detailed treatments such as a self-consistent hydrodynamic code. Such studies will be particularly important once the rate uncertainty has been further reduced through measurements of the unknown resonance strengths in Table II.

IV. CONCLUSIONS

We have improved calculations of the thermonuclear $^{33}$S(p,γ)$^{34}$Cl rate in ONe novae by clarifying the level structure of $^{34}$Cl within 1 MeV of the proton threshold. The $^{33}$S(p,γ)$^{34}$Cl reaction is a key pathway that influences the late stages of the abundance flow for nova explosions on the most massive ONe white dwarfs. Uncertainties in this rate affect nova nucleosynthesis predictions for elements above sulphur [19], and may also determine whether or not the presence of large amounts of $^{33}$S relative to solar is a viable diagnostic for the nova paternity of presolar grains. Possibilities for detection of β-delayed γ-rays from $^{34m}$Cl ($t_{1/2}$ = 32 min) – in nova explosions produced only by $^{33}$S(p,γ) – also depend upon this rate.

Our new $^{33}$S(p,γ) rate is as much as 26 times larger than previous calculations at relevant temperatures in novae (~0.3 GK). The rate is still highly uncertain, though, (by at least an order of magnitude at T = 0.3 GK) because of the experimentally-unknown resonance strengths of $^{34}$Cl states within ~600 keV of the $^{33}$S+p threshold. This prevents strong conclusions to be drawn about the effects of our results on nucleosynthesis in ONe novae. With calculations from Iliadis et al. (2002) [19] and José et al. (2001) [8], however, we can make some simple estimates. We find reductions by as much as a factor of 12 in predictions of $^{33}$S production in ONe novae, implying a $^{32}$S/$^{33}$S ratio in nova ejecta of ~ 1160. For comparison, $^{12}$C/$^{13}$C and $^{14}$N/$^{15}$N isotopic ratios predicted in ONe nova ejecta are both ~ 1 [7]. The predicted $^{32}$S/$^{33}$S ratio represents a large isotopic deficiency of $^{33}$S with respect to $^{32}$S in comparison to solar ($^{32}$S/$^{33}$S = 127). We stress, however, that our estimates must be taken with caution because of the large rate uncertainty: for example, using our (very conservative) upper and lower limits for the rate at T = 0.3 GK, the $^{32}$S/$^{33}$S ratio could vary by an order of magnitude. Finally, it is unlikely that our new rate will dramatically change the amount of $^{34m}$Cl expected from ONe novae, supporting results from José et al. (2001) on the faint prospect of detecting γ-rays following the decay of any $^{34m}$Cl produced in nova explosions [8].

Future experimental work should focus on measuring the unknown (p,γ) resonance strengths of the proton-threshold states in $^{34}$Cl (especially those of the new resonances at $E_R$ = 281, 301 and 342 keV, which may dominate the $^{33}$S(p,γ) rate at T ~ 0.3 GK). Suitable experiments could involve measuring $^{33}$S(p,γ) directly with a high-intensity proton beam and an enriched $^{33}$S target (this could also be performed in inverse kinematics [46]), and/or measuring the $^{33}$S($^{3}$He,d)$^{34}$Cl reaction to determine $J^\pi$ values and proton spectroscopic factors for these states (which could then be used to determine proton partial widths $\Gamma_p$). Such measurements are currently being planned and will certainly help to provide a better evaluation of the $^{33}$S(p,γ) rate. Other transfer-reaction studies populating proton-threshold states in $^{34}$Cl would also be useful to help constrain the $J^\pi$ values. Measurement of γ-ray branching ratios of these $^{34}$Cl states, as well as of reactions relevant to the destruction of $^{34}$Cl, would help to better define prospects for the observation of β-delayed γ-rays from $^{34m}$Cl. Finally, we wholeheartedly encourage more extensive measurements of sulphur isotopic ratios in presolar grains.

Acknowledgments

It is a pleasure to thank the crew of the MLL tandem accelerator. We also greatly appreciate comments and suggestions from J. José and C. Iliadis on an early version of this manuscript.

This work was supported by the DFG cluster of excellence "Origin and Structure of the Universe" (www.universe-cluster.de).

Table I: Level structure of 34Cl between 4.92 < $E_x$ < 5.98 MeV.  Levels from the present study are shown with uncertainties due to statistics and reproducibility.  In addition, a ± 3 keV uncertainty arises from systematic uncertainties (see text); the good agreement between energies from the present work and those from past work indicate that the systematic uncertainty may be overestimated.  33S(p,γ)34Cl resonance energies $E_R$ are calculated with weighted averages between past and present work when possible, using Q = 5142.75(12) keV [16].

| $E_x$ (keV) [20] | $E_x$ (keV) Present work | $J^\pi$ [20] | $E_R$ (keV) | (ωγ) (meV) [20] |
|---|---|---|---|---|
| | 4925(2) | | | |
| 4941.9(4) | 4942(2) | (1,2)+ | | |
| 4957.7(10) | 4957(2) | (2-4)+ | | |
| 4971(11) | 4981(2) | (1,2)+ | | |
| 4995.6(3) | 4995(2) | (1,2)+ | | |
| 5010(13) | 5010(2) | 0+ | | |
| | 5061(2) | | | |
| | 5093(2) | | | |
| | 5154(2) | | 11 | |
| 5171.6(3) | 5173(2) | 4 | 29 | |
| | 5233(2) | | 90 | |
| | 5263(2) | | 120 | |
| 5315.0(3) | | 7+ | 172 | |
| | 5326(2) | | 183 | |
| | 5357(2) | | 214 | |
| 5386.8(15) | (5388) [a] | (4-6-) | 244 | |
| | 5424(2) | | 281 | |
| | 5444(2) | | 301 | |
| | 5485(2) | | 342 | |
| 5540.8(11) | 5545(2) | (4,5-) | 399 | |
| 5576(1) | 5572(2) | 3 | 432 | 50(13) |
| | 5606(2) | | 463 | |
| 5635.0(3) | 5635(2) | (1,2+) | 492 | 88(25) |
| 5672(1) | (5671) [a] | (1,2)+ | 529 | 88(38) |
| | (5705(4)) | | 562 | |
| 5762(1) | (5761) [a] | (1,2 +) | 619 | 8(4) |
| 5785(1) | 5782(2) | (1+ - 3-) | 642 | 50(25) |
| 5805(1) | 5806(2) | 2- | 662 | 50(25) |
| 5852.1(3) | 5852(2) | (2,3)- | 709 | 63(25) |
| | 5868(2) | | 725 | |
| 5896(1) | 5898(2) | 2 | 754 | 63(25) |

| | 5917(2) | | 774 | |
|---------|---------|-----|-----|---------|
| 5940(1) | 5941(2) | 2+  | 797 | 100(25) |
| | 5980(2) | | 837 | |

( ) = observed at $\theta_{lab}$ = 15° only

a = tentative observation (low statistics)

Table II: $^{33}$S(p,γ)$^{34}$Cl resonance parameters assumed for known $^{34}$Cl states with unknown (p,γ) resonance strengths (see Table I). See text for details. These were used in conjunction with states for which measurements of (p,γ) resonance strengths exist (see Table I and [20]) to calculate an *upper limit* for the $^{33}$S(p,γ)$^{34}$Cl rate – see the upper dashed grey line in Fig. 2a.

| $E_x$ (keV) | $E_R$ (keV) | $J^\pi$ | l | $C^2S$ | $\theta^2_{s.p.}$ | $\Gamma_p$(meV) | (ωγ) (meV) |
|---|---|---|---|---|---|---|---|
| 5154 | 11  | 2+ | 0 | 1 | 0.55 | 2.3 x 10$^{-52}$ | 1.4 x 10$^{-49}$ |
| 5172 | 29  | 4+ | 2 | 1 | 0.30 | 5.5 x 10$^{-29}$ | 6.2 x 10$^{-29}$ |
| 5233 | 90  | 2+ | 0 | 1 | 0.55 | 1.2 x 10$^{-9}$ | 7.4 x 10$^{-10}$ |
| 5263 | 120 | 2+ | 0 | 1 | 0.55 | 1.2 x 10$^{-6}$ | 7.8 x 10$^{-7}$ |
| 5315 | 172 | 7+ | 6 | 1 | 0.1  | 2.0 x 10$^{-14}$ | 3.8 x 10$^{-14}$ |
| 5326 | 183 | 2+ | 0 | 1 | 0.55 | 6.2 x 10$^{-3}$ | 3.9 x 10$^{-3}$ |
| 5357 | 214 | 2+ | 0 | 1 | 0.55 | 9.4 x 10$^{-2}$ | 5.9 x 10$^{-2}$ |
| 5387 | 244 | 4+ | 2 | 1 | 0.31 | 7.2 x 10$^{-3}$ | 8.1 x 10$^{-3}$ |
| 5424 | 281 | 2+ | 0 | 1 | 0.55 | 6.7 | 4.2 |
| 5444 | 301 | 2+ | 0 | 1 | 0.54 | 18 | 11 |
| 5485 | 342 | 2+ | 0 | 1 | 0.54 | 1.0 x 10$^2$ | 63 |
| 5542 | 399 | 4+ | 2 | 1 | 0.31 | 7.6 | 8.6 |
| 5606 | 463 | 2+ | 0 | 1 | 0.54 | 4.0 x 10$^3$ | 2.5 x 10$^3$ |
| 5868 | 725 | 2+ | 0 | 1 | 0.54 | 3.5 x 10$^5$ | 2.2 x 10$^5$ |
| 5917 | 774 | 2+ | 0 | 1 | 0.53 | 6.0 x 10$^5$ | 3.8 x 10$^5$ |
| 5980 | 837 | 2+ | 0 | 1 | 0.53 | 7.9 x 10$^5$ | 4.9 x 10$^5$ |

FIGURE CAPTIONS

FIG. 1: (Color online) Focal-plane triton spectra (black) from the $^{34}$S($^{3}$He,t)$^{34}$Cl reaction at 25 MeV and (a) $\theta_{lab}$ = 15°, (b) $\theta_{lab}$ = 25 deg. Level energies are in keV. These spectra are shown adjusted relative to one another to roughly preserve the scale for $^{34}$Cl excitation energies. Also shown are normalized triton spectra obtained with a natural Ag target (on a C backing) to illustrate the background (gray; red online).

FIG. 2: (a) Thermonuclear $^{33}$S(p,$\gamma$)$^{34}$Cl reaction rate over typical nova T$_{peak}$ – see text and Table II. The solid black line is the geometric mean of the upper and lower dashed grey lines. (b) Ratio of the nominal $^{33}$S(p,$\gamma$)$^{34}$Cl rate from the present work (solid black line in (a)) to that of Iliadis et al. (2001) ([18], filled grey squares), and to that of José et al. (2001) ([8], solid black line).

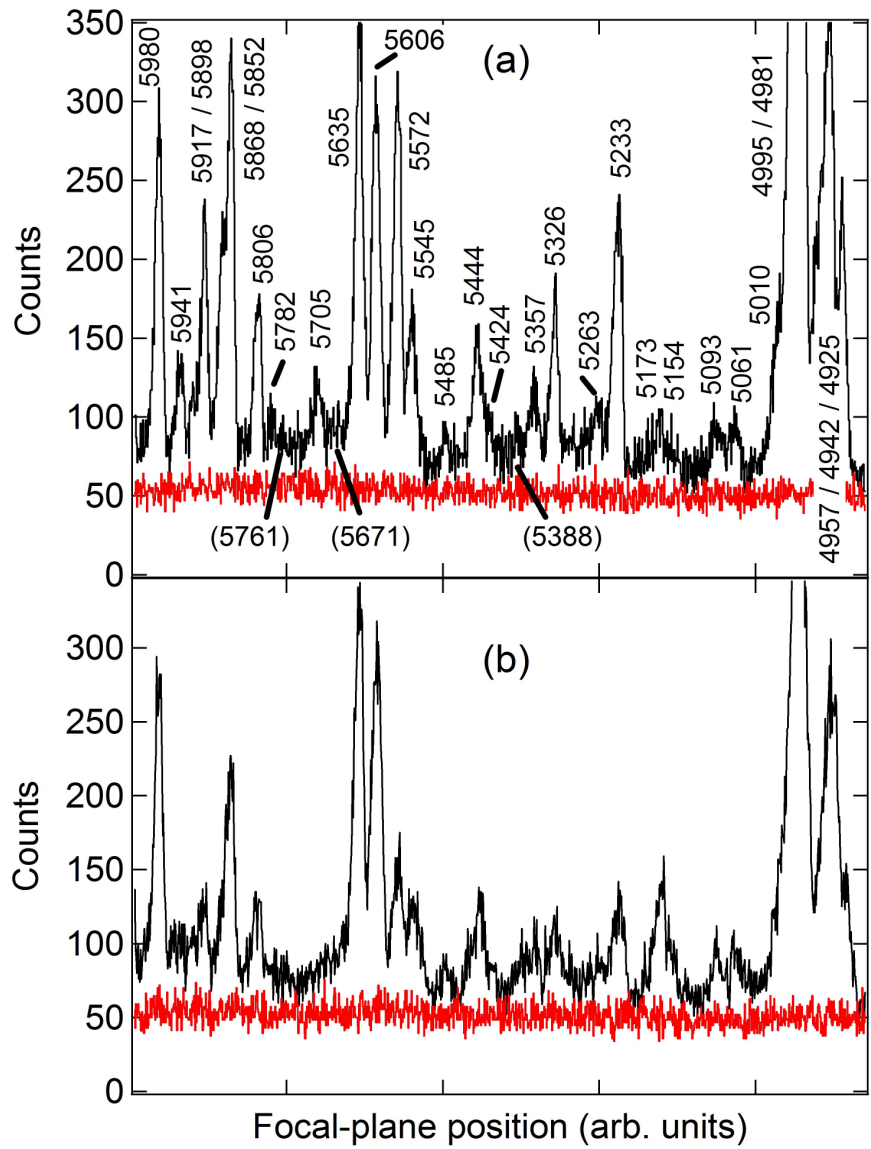

FIG. 1

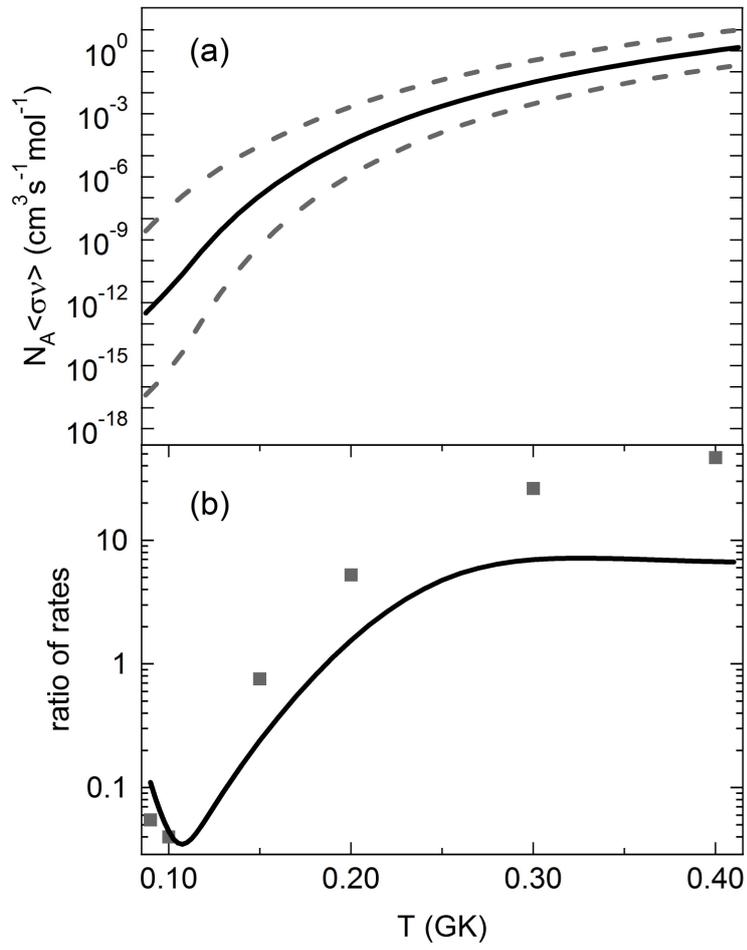

FIG. 2